# Experiences from Large-Scale Model Checking: Verification of a Vehicle Control System

Jonas Fritzsch
University of Stuttgart
Institute of Software Engineering
Stuttgart, Germany
jonas.fritzsch@iste.uni-stuttgart.de

Tobias Schmid
BMW AG
Munich, Germany
tobias.ts.schmid@bmw.de

Stefan Wagner
University of Stuttgart
Institute of Software Engineering
Stuttgart, Germany
stefan.wagner@iste.uni-stuttgart.de

*Abstract*—In the age of autonomously driving vehicles, functionality and complexity of embedded systems are increasing tremendously. Safety aspects become more important and require such systems to operate with the highest possible level of fault tolerance. Simulation and systematic testing techniques have reached their limits in this regard. Here, formal verification as a long established technique can be an appropriate complement. However, the necessary preparatory work like adequately modeling a system and specifying properties in temporal logic are anything but trivial. In this paper, we report on our experiences applying model checking to verify the arbitration logic of a Vehicle Control System. We balance pros and cons of different model checking techniques and tools, and reason about our choice of the symbolic model checker NuSMV. We describe the process of modeling the architecture, resulting in ~1500 LOC, 69 state variables and 38 LTL constraints. To handle this large-scale model, we automate and optimize the model checking procedure for use on multi-core CPUs and employ Bounded Model Checking to avoid the state explosion problem. We share our lessons learned and provide valuable insights for architects, developers, and test engineers involved in this highly present topic.

*Index Terms*—Formal Verification, Model Checking, Experience Report, NuSMV, Driving Automation, Vehicle Control System, Bounded Model Checking, NuSMV

## I. INTRODUCTION

Formal verification is of particular importance in addressing safety-critical aspects of software-intensive systems, with an example being the current state of the automotive industry. The recent trend in this domain is characterized by advancements towards autonomously driving vehicles. A higher level of automation inevitably increases the complexity of involved embedded systems. In this regard, *"it is increasingly advisable to formally verify their adherence to safety properties to prevent loss of life"* [1]. IBM's recently published study *"Automotive 2030"* even predicts that in a decade's time, safety credentials will become the brand differentiator for autonomous vehicles [2]. Hence, new ways and approaches are required to ensure reliability and safety in accordance to established standards. In this context, functional safety as a *"part of the overall safety that depends on a system or equipment operating correctly in response to its inputs"* [3] is associated with the involved electrical, electronic, and programmable systems. Car manufacturers thus need to *"prove that the designed system fulfills the functional safety requirements according to ISO 26262"* [4] which is mandatory for the certification of automated driving systems [5].

Traditional testing and simulation techniques, or failure mode and effects analysis (FMEA) alone reach their limits in this regard. An exhaustive verification with limited resources and time is difficult to achieve or even impossible. Hence, such established approaches need to be complemented by formal verification techniques to thoroughly validate safety critical embedded systems. To this effect, model checking has been widely used in similar contexts [6], [7], [8], moreover *"all major microprocessor manufacturers use model checking techniques to verify their processor designs"* [6].

In this paper, we report on our experiences gathered during model checking a Vehicle Control System. Next to *"choosing an appropriate level of abstraction for the model"* [9] and adequately formalizing specifications, the size and complexity of our resulting model required particular efforts. Increasing complexity also limits the capabilities of formal verification methods, with the *state explosion problem* being *"the main obstacle that model checking faces"* [9]. The problem is exacerbated by systems with an inherent complicated timing behavior. A such, we tackle the *state explosion problem* using Bounded Model Checking [10]. We employ the symbolic model checker NuSMV [11] which has been widely used in similar contexts and proven to be suitable for large scale models [12], [13]. Besides, we automate the checking procedure for use on multi-core processors and share our lessons learned along the verification project.

As per our knowledge, there have been no similar experiences shared in scientific literature so far on model checking a Vehicle Control System with the aid of NuSMV. Verifying systems of high complexity requires to choose the right tools and techniques. The absence of guidelines in this regard has been pointed out by Choi already [13]. As such, we are confident that our contribution can pro-vide valuable insights for architects, developers, and test engineers that develop systems with similar requirements on functional safety and fault tolerance.

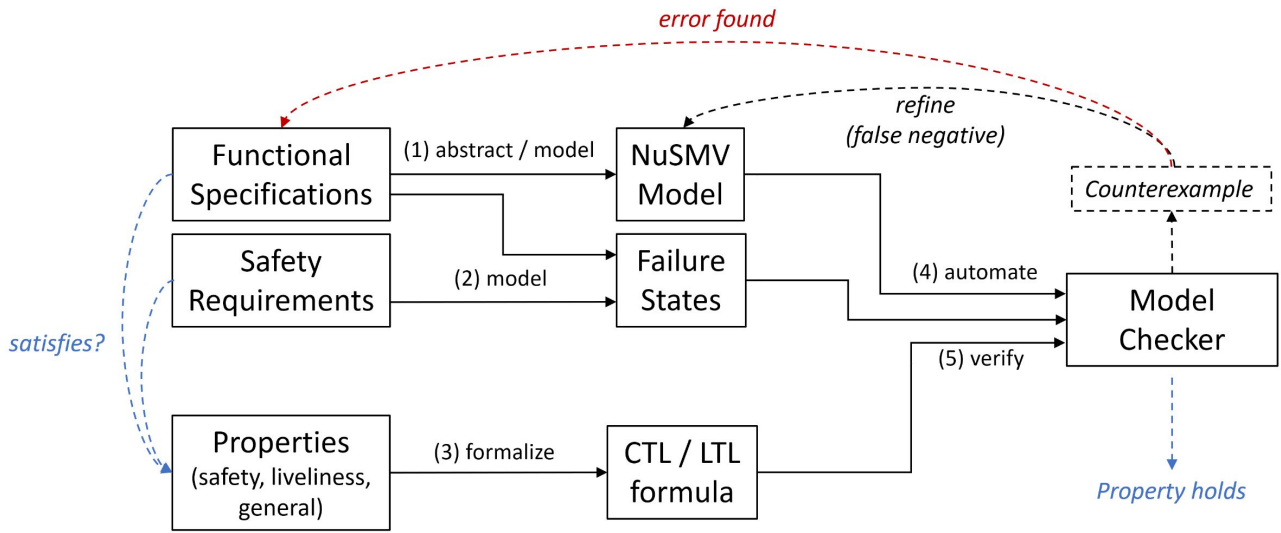

Fig. 1: Formalization and Verification Process

The remainder of this paper is structured as follows: Section 2 describes our methodology, while section 3 discusses related work. In section 4, we introduce our case study by describing the architecture of the Vehicle Control System to be verified. Section 5 is devoted to modeling the presented system and formalizing specifications. Section 6 will outline our approach to automate and optimize the model checking procedure. We conclude with a discussion of limitations in section 7 and a with a brief summary in section 8. In sections 5 and 6 we point out our lessons learned along the project.

## II. Methodology

We report on our experiences gathered during the formal verification of a Vehicle Control System using the NuSMV symbolic model checker. Figure 1 depicts the process we followed and shows involved artifacts. Starting from functional specifications and a set of related properties, we describe the creation of a formal model and the necessary abstractions we made in this process (1). Based on given safety requirements, a number of failure states needed to be modeled (2). Next, we formalize the relevant properties and report experiences when deriving CTL/LTL formula (3). The complexity of the system requires an automation of the model checking procedure. We present our approach to partition the checking by failure state combinations and batch processing the model checking (4). In the course of it, we optimize the verification procedure by enabling NuSMV to utilize multiple processor cores (5). Besides, we discuss synchronous and asynchronous timing behaviors and use of Bounded Model Checking to avoid the *state explosion problem*. When appropriate, we highlight aspects that we perceived important and share our lessons learned.

## III. Related Work

In this section, we briefly discuss efforts made using model checking in the automotive domain as well as similar documented experiences from other domains, focusing on large scale models with high complexity and the model checker NuSMV.

Köbl and Leue present a method to assess functional safety of architectures for Automated Driving Systems (ADS) according to the ISO 26262 standard. Their model-based approach relies on Systems Modeling Language (SysML) and QuantUM, a tool for model and causality checking that is *"well suited to deal with concurrency induced non-determinism"* [14]. Compared to our work, their approach models the system on a higher level of abstraction. It can be seen as an alternative way for verifying such architectures using a different tool chain. Similarly, Abdulkhaleq and Wagner propose an approach to model checking safety-critical systems using the model checker SPIN [15]. They verify software safety requirements and illustrate their approach by means of an adaptive cruise control system. They formulate specifications in Linear Temporal Logic (LTL) that are derived from a STPA safety analysis.

Nyberg et al. report on their experiences in applying formal verification techniques to embedded, safety-critical code in the automotive domain [16]. From gathered knowledge over eight years, they describe six industrial case studies conducted at Scania. The authors emphasize the immanent difficulty of formalizing complex system properties: *"the high-level engineering artifacts that we were supposed to verify against were highly ambiguous"* [16]. A higher level of abstraction would exacerbate the situation. They summarize a number of additional obstacles and suggest according remedies in form of a roadmap.

Other large-scale model checking experiences stem e.g.

from the aviation domain. Boniol et al. report on using model checking for verifying a landing gear control system of an aircraft. Among other model checkers, they employ Uppaal and NuSMV. Their goal was to evaluate the usability of different model checking tools in an industrial context. NuSMV proved to be the most efficient one and *"the only tool to succeed in all steps of the experimentation, even the most complex ones"* [17]. Their modeling approach for a discrete real-time system is very similar to our case. They also take different precautions to reduce complexity of the model and to avoid the *state explosion problem*.

Choi reports on experiences with model checking flight guidance systems [13]. Aiming to assess the capabilities of two contrasting techniques, they employ explicit model checking using SPIN and symbolic model checking using NuSMV. They note that *"the choice between symbolic and explicit model checking [...] is quite unclear, with only limited comparative arguments"* [13]. Each technique would have its own way to tackle complexity and avoid the *state explosion problem*. It comes up to weighing exhaustiveness against run time, eventually no model checking tool was clearly superior over the other.

The most extensive work we found was reported by Lahtinen [12]. He reports on verifying the fault-tolerance of a nuclear power plant using model checking with NuSMV. As in autonomous driving, such plants rely on redundant subsystems to achieve a certain level of fault-tolerance. Their work targets the exhaustive verification based on a *"single model that covers the entire plant"* [12]. Consequently, they face the issue of a very large model. They apply certain abstraction techniques to handle the resulting complexity which significantly increased when considering timing behavior for modules.

While advances in model checking go back some time, *"formal verification is not yet state-of-practice in the automotive industry"* [16]. The difficulty of formalizing complex system properties may hinder its use [16], as well as the difficulty of choosing the right technique and tool [13]. To this end, we describe experiences and lessons learned that aim to mitigate this issue and facilitate the adoption of model checking techniques in industry.

## IV. Vehicle Control System

In this section, we introduce the system under verification by providing more context on the architectural requirements of automated driving and subsequently outlining the high-level architecture of our Vehicle Control System.

### A. Automated Driving

In automated driving, there are five levels of automation ranging from no driving automation (level 0) to full driving automation (level 5) [18]. While in level 2 (partial automation) the driver is still responsible for supervising the driving task, level 3 extends this to conditional automation where *"the driver is not 'actively' required to monitor the environment, so called 'hands and eyes off'"* [4]. The driver is required to re-assume active driving only when notified by the vehicle, i.e. the automated driving system. This shift to level 3 requires to extend a fail-safe system design (level 2) to a fail-operational system design (level 3 and above). A fail-safe system needs to only shut down the driver assistance functionality in a fail-safe way on occurrence of an error. By contrast, a fail-operational system needs to implement degradation mechanisms, as the driver may not take over control again on request. A limited time of degraded operation needs to be bridged, e.g. until triggering a safety brake of the vehicle. A malfunction of the driving system even for cascading failures needs to be avoided as well. The assurance of such controlled operation in the presence of failures poses new challenges on test and verification strategies [19].

### B. Vehicle Control System Architecture

A fail-operational conform system design requires redundancy of the involved subsystems and Electronic Control Units (ECU), such as braking and steering. Kron proposes such an architecture to enable conditional (level 3) automation [4]. It is composed of separate subsystems for observing, interpreting, and acting tasks. The environment is observed by various sensors and interpreted to an environmental model. This model is enriched by the inclusion of cloud data that is further processed involving road and weather conditions or even information from other nearby vehicles. A resulting calculated trajectory is then used to control different types of actuators, such as engine, brakes and steering systems. Such a vehicle *"is able to brake and steer in order to keep following a given driving trajectory for the period of time the automated driving function remains active"* [4]. To do so, the system design must be able to tolerate hardware or software faults of included subsystems, and *"ensure the availability of the motion control system including its actuators [...] for a certain amount of time"* [4]. An architecture capable of fulfilling these requirements is characterized by an increased complexity.

Kohn et al. state that *"electronic components without mechanical fallback require a fail-operational implementation to guarantee a correct safety-behavior"* which can be *"achieved either by design diversity or redundancy"* [20]. In their work-in-progress they present an overview of such architectures for fail-operational automotive systems. The architecture described in our study is based on a "2-out-of-2 DFS Architecture" accordingly. As such, it consists of two independent subsystems, a primary control system and a backup system. Avoiding full redundancy minimizes the needed components, but requires an arbitration logic that allows to dynamically switch to different fallback operation modes next to the normal mode for faultless operation. Figure 2 shows a simplified schematic representation of this architecture.

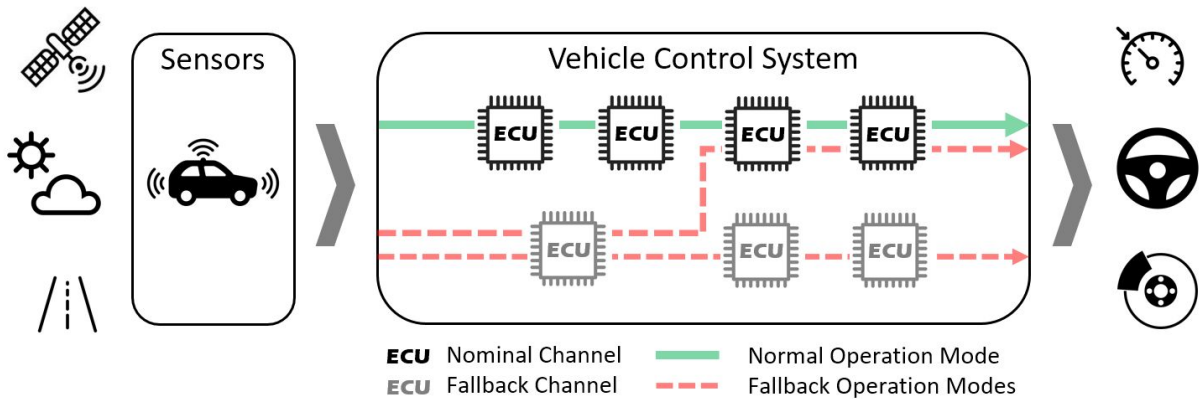

Fig. 2: Vehicle Control System based on a 2-out-of-2 DFS Architecture [20]. Icons from [21]

As errors may occur in any of the involved components and subsystems, *"sophisticated error detection and system switching algorithms"* [4] need to be implemented. In our architecture, an arbitration logic controls the functional interaction between all involved ECUs and thus determines the active operation mode. Next to the normal mode for faultless operation, one of three fallback channels can be activated, depending on the faulty ECU or subsystem. Subsystem, here, refers to different vehicle bus systems that connect a subset of the ECUs as well as the redundant power supply. Such a redundant system design results in several possible combinations of intertwining ECUs and subsystems that need to be controlled by arbitration logic.

### C. Verification Goals and Specifications

The goal of this verification is prove that our architecture resists any kind of single and double failures. Such failures can be any of the following kinds: ECU malfunctions, communication failures, and power supply outages. Failure resistance in this regard means that a predetermined operation mode needs to be activated by the arbitration logic within a certain range of time after failure occurrence. The goal of a formal verification is in line with the requirements of an ISO 26262 [22] compliance. Simulation and systematic testing have reached their limits in this regard. A full coverage, i.e. exhaustive verification would be infeasible by their means, given the complexity of the system. Hence, formal verification was chosen to complement these existing techniques. While formal verification requires to abstract from the hardware and software level, an elaborate modeling of the system behavior is required in advance. Such an abstraction step, however, carries the risk of missing certain kinds of failures, introducing new failures (resulting in false negatives), or even causing unintended malfunction. Another goal is therefore a critical reflection of benefits, limitations, and credibility of formal verification techniques and their results.

The system's architecture was provided in form of a functional specifications document that is used for e.g. third-party component manufacturers of ECUs. As well, safety requirements derived from the ISO26262 standard were provided. They describe a set of possible failure states and combinations that need to be considered. These failure combinations were given in form of a matrix that states the targeted operation mode resulting from each double failure combination.

### V. Modeling

In this section, we describe the process of modeling the above described architecture and formulating constraints in temporal logic. As a first step, we outline our considerations regarding the selection of a model checking technique and tool. Next, we transform the functional requirements into the model checker's description language. In connection with it, we model the failure states of the system architecture as given by functional and safety requirements. In the last step, we formulate safety, liveliness, and general properties.

### A. Tool Selection

There exists a variety of tools for model checking [23]. The following tool requirements were prescribed and allowed us to narrow down the choices. The tool should have an established user base and proven its capabilities for checking complex models. Open-source software that can be used on multiple platforms was preferred. In this regard, it should as well be suitable for a tool qualification according to ISO 26262. Regarding the expressiveness of its specification language, a coverage of LTL and CTL temporal logic was desired. Likewise, it should be capable of modeling concurrent systems with asynchronous behavior. As a result, we shortlisted the following three model checkers: SPIN [24], UPPAAL [25], and NuSMV [11].

UPPAAL had been used in some related work already [26], [8], [17] and distinguishes itself from the other two by the possibility to specify the model using a graphical editor. However, we recognized limits regarding its expressiveness. While a graphical editor might be convenient in the beginning, bulk changes in complex models later on might be time consuming. As well, UPPAAL seemed not as widely used as the others, and commercial use was not free of charge. Instead, both SPIN and NuSMV met nearly all of the above stated requirements. The two model checkers follow a contrary strategy. SPIN relies on *"explicit representation of the transition system associated to a model specification"* [27], while NuSMV represents the transition system as a Boolean formula, so called Binary Decision Diagrams (BDD).

As we expect to encounter the *state explosion problem* due to the complexity of our system, strategies to mitigate this issue were of equal importance. SPIN as an explicit state model checker *"draws its main power from partial order reduction techniques"* [10] in this regard. NuSMV, on the contrary, employs Bounded Model Checking (BMC) based on a Satisfiability Solver (SAT) [28]. Both strategies have their benefits and limitations. The implementation of partial order reduction in SPIN allows, next to exhaustive checking, hash-compact and bit-state hashing [13]. Both options can leave some states unexplored and thus miss to detect counterexamples. In Bounded Model Checking, as used by NuSMV, *"a transition system and a property are jointly unwound for a given number k of steps"* [28]. That means, up to a certain bound given by k, the model is checked exhaustively. Choi reports on his experiences using both SPIN and NuSMV comparatively for model checking flight guidance systems [13]. Due to the complexity of their system, the use of an exhaustive checking algorithm was not applicable either. They could not determine a clearly superior approach, the choice between explicit state (SPIN) and symbolic (NuSMV) model checking could not be reasonably argued [13].

Gaining these insights, we started a prototypical modeling using both SPIN and NuSMV. We chose NuSMV for our subsequent implementation due to several documented successful experiences [12], [13], [17]. Besides, we valued NuSMV's greater expressiveness which allows specifications in LTL and CTL, including its real-time extension RTCTL. It should turn out that we were finally able to exclusively use LTL formula for our specifications. However, in the first iterations of the project we experimented with CTL and RTCTL that often allowed more compact and understandable formulations.

### B. Modeling the System Architecture

We modeled each of the ECUs as a module in NuSMV's description language. The states, possible transitions between them, and related state variables were detailed in the functional specifications document and could be translated directly. Each of the ECUs can be represented by a finite state machine, as Figure 3 exemplarily shows. In the main module of the model, each ECU module is instantiated. Every ECU module contains an internal state variable representing its current state. The possible values of this internal state variable are given by the valid states of this ECU as shown in Figure 3, enhanced by failure states. This variable is getting recalculated at every execution step, while the new state is based on the arguments passed to this ECU module. To exchange data, subsets of ECUs are connected over different bus systems. This communication is realized via parameter passing between the ECU modules. Depending on the ECU, the parameter list contains between 5 and 11 parameters. Hence, several expressions need to be evaluated in a certain order to determine the module's next state. The first expression evaluated to true determines the following state of the ECU. Listing 1 shows the basic structure of a module in NuSMV. Module `M_ECU1` calculates its state based on the current states of ECU2 and ECU3. The module's internal state variable `S_ECU1` can be used as input for other modules likewise. External failure states, here `FailA` and `FailB`, also affect the module's state calculation.

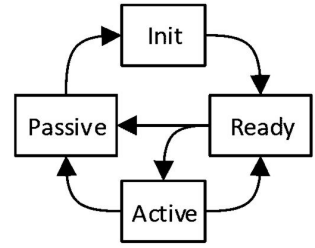

Fig. 3: Typical Transition Diagram of an ECU

Listing 1: Modeling an ECU in NuSMV

```
MODULE M_ECU1 (S_ECU2, S_ECU3, FailA, FailB)
VAR
    S_ECU1 : {Init, Ready, Active, Passive};
ASSIGN
    init (ECU1) := Init;
    next (ECU1) :=
        case
            FailA & FailB: Passive;
            S_ECU1 = Init & !FailA : Ready;
            S_ECU1 = Ready & S_ECU2 = Ready
                & !FailA & !FailB : Active;
            ...
            TRUE: S_ECU1;
        esac;
```

As the verification of a discrete time model proceeds stepwise, a step is the finest possible temporal granularity. Each step entails a re-calculation of all state variables in the model with an instantaneous propagation. In line with the functional specifications, we decided that each step correlates to a processing cycle of 10ms in the physical system. This is a first major abstraction we introduced. A challenge in this respect was the implementation of a required debouncing behavior in case of a communication malfunction, e.g. when a signal is not properly transmitted from one ECU to another. As per requirements, the receiver should then assume the last communicated state of the sender for three cycles (30ms), before assuming a communication failure or failure state of the sender. This behavior could not be modeled straight forward using parameter passing. We therefore added a separate bus module that implements the debouncing logic and acts as a proxy for all communication between the ECU modules. The bus module abstracts all communication taking place over different vehicle bus systems. As such, our NuSMV model has a module structure as depicted in Figure 4.

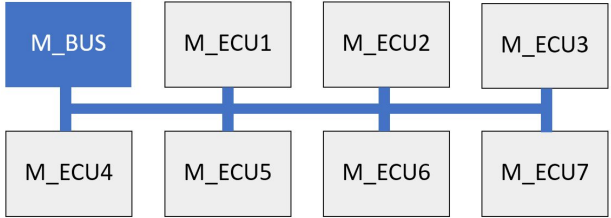

Fig. 4: Abstraction of Communication in the NuSMV Model

Next to the instantiated ECU and bus modules, the main module also contained Boolean state variables for the operation modes, indicating the active mode. The active operation mode is calculated based on the current states of all ECUs. As such, these variables are an essential part of the arbitration logic and used in formulating constraints in temporal logic later on. At this point in time, the model comprised 9 modules including the main module and 44 state variables. As we had introduced no failure states yet, the model checker could execute the model deterministically. This allowed us to verify the run-up phase of the modeled system which results in the normal operation mode being active after 15 steps.

> **Lesson Learned: Global Constants**
>
> The syntax of NuSMV does not allow the definition of global constants. While a macro using the `DEFINE` keyword can be used alternatively, it is only allowed within modules. Macros do not introduce a new variable and thus do not increase the state space. Defining them in the main module would require to explicitely pass them to other modules, which is not a very elegant solution. As a workaround, we created a `M_GLOBAL` module that contained all macros. This module was instantiated in the main module and its instance variable could then be passed to all other modules, allowing them to access several global definitions through prefixing the passed instance variable.

> **Lesson Learned: Measuring Elapsed Time**
>
> At several places we needed to measure the number of elapsed time to trigger a certain transition, e.g. when ensuring the switch of an operation mode within a certain number of steps. For such purposes, we implemented a timer according to the following pattern:
>
> ```
>     Timer: 0..Global.T1_MAX;
>     init (Timer) := 0;
>     next (Timer) :=
>     case
>         <Condition_1> : 0;
>         Timer = Global.T1_MAX : Timer;
>         <Condition_2> : Timer + 1;
>         TRUE : Timer;
>     esac;
> ```
>
> This Timer can count up from 0 to `T1_MAX`. `Condition_1` would reset the timer, `Condition_2` would increase it, and in any other case the timer would be paused. Here, we used `T1_MAX` as a macro that was defined the `Global` module.

### C. Modeling the Failure States

In the next phase, we modeled the failure states as given by functional specifications and safety requirements. Next to a failure state of each ECU that indicates its outage, the communication between ECUs can malfunction in several ways. Lastly, failures caused by the outage of a power supply needed to be considered. The ECU and power supply failure states required 10 Boolean variables, whereas modeling the communication failures turned out to be more extensive. While the outage of each of the three involved bus systems could be modeled as a single failure state as well, the entirety of point-to-point communication paths needed a separate handling. This resulted in 26 additional communication-related failure states, summing up to 39 basic failure states. These failure states were implemented as Boolean state variables and arrays in the bus and main modules. The complex arbitration logic required an additional calculation of several composite failure states that trace back to a subset of the above mentioned 39 basic failure states. Consequently, these composite failure states could be evaluated only delayed by one step. This necessary abstraction was owed the discrete time logic used.

As a result of modeling failure states, we implemented 42 Boolean states variables that indicate a failure at any point in time during operation. The safety requirements stated that the arbitration logic needs to resist the occurrence of any single failure or double failure combination. While a single failure could occur at any point in time and also remain active for any period of time (see Figure 5), failure combinations require further considerations: Is the order of occurrence important? Are sequential occurrences and overlaps treated equally? Does the length of the overlap matter? The functional specifications did not exhaustively discuss these possibilities. Hence, we agreed

on the following assumptions affecting single and double failures:

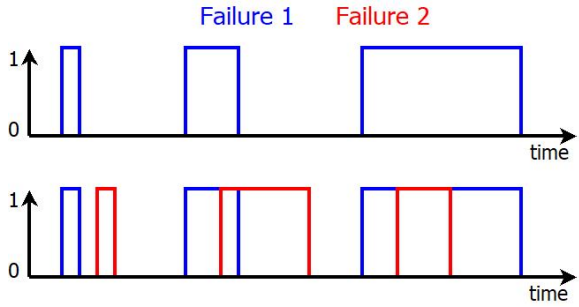

Fig. 5: Occurrences of Single and Double Failures

1) Each failure occurs only once in a single verification run.
2) Each failure can occur at any point in time and can remain present for any period of time.
3) For double failures, the order of occurrence matters, as the target operation mode depends on this order.
4) For double failures, sequences and overlaps are treated equally, i.e. both any or no overlap are possible.

*D. Formulation of Constraints*

The specifications stated several properties that needed to be verified. These properties can be divided into general, liveliness, and safety-related properties. General and liveliness properties ensure activation, deactivation, and deadlock-free operation of the system. Other properties of this type addressed the valid order of operation modes or a switch between them within a certain number of steps. All those properties had to be verified in failure-free operation, as well as in combination with possible failure states.

The safety-related properties were derived from *safety goals* which had been defined in accordance to the ISO 26262 standard. These safety-related properties were covered in our model by an exhaustive verification of the 42 possible failure states and combinations. Each combination requires the arbitration logic to activate a predefined target operation mode, as far as the combination is not fatal as e.g. the outage of primary and backup power supplies at the same time.

Initially, we formulated the constraints in LTL, CTL, or RTCTL, depending on the suitability. NuSMV accepts specifications in all three temporal logic dialects. Especially time-dependent specifications like *"switch to a state within a certain number of cycles and keep this state for another number of cycles"* could be conveniently converted to RCTL, a real-time extension of CTL. At the end, we had formulated more than 30 constraints that were partly complex expressions including evaluations of up to 20 state variables. However, not every constraint had to be verified for all failure combinations. A subset of these constraints was relevant for verifying failure-free, single failure, and double failure scenarios respectively.

**Lesson Learned: Asynchronous Timing**

In our discrete time model, all state machines execute in parallel. Such a synchronous behavior implies a major abstraction of the physical system. As communication between ECUs involves different bus systems, executions are actually interleaved. NuSMV offers capabilities for modeling asynchronous timings between modules as well. Unfortunately, the functionality has been deprecated from version 2.5.0 upwards. The documentation briefly states that *"modeling of asynchronous processes will have to be resolved at higher level"* [29]. Consulting the NuSMV FAQs revealed that is was removed due to inconsistencies and additional complexity.

Nonetheless, we experimented attaching the deprecated keyword `PROCESS` to our modules. It causes the model checker to execute a step *"by non-deterministically choosing a process, then executing all of the assignment statements in that process in parallel"* [29]. Intentionally, it was the behavior we were looking for, but our tests showed that it does not appropriately reflect the system's behavior. The model checker considers any idle period between consecutive executions of the same process as feasible. Consequently, plenty of counterexamples were generated that turned out to be false negatives.

The actual behavior we were aiming for is depicted in Figure 6. The default synchronous timing is shown in (a) where modules M1 and M2 are always executed at the same time. Illustration (b) shows an interleaved execution of modules M1 and M2. Hence, module M1 can pass arguments to M2 and does not execute again until M2 returns its results. Variant (c) extends this interleaved execution by another step.

We in fact aimed for the model checker to consider a one-step interleaved execution in any order when performing the verification. However, such a behavior would need to be implemented *"at higher level"* [29], i.e. in the model itself. Such a pause state per module could be realized by introducing a dedicated state variable for each module. More difficult to answer is the question, how many steps a module should be able to pause. Next to the additional complexity, this behavior would multiply the time needed for a verification run. Hence, we postponed pursuing this idea further, but it may be a viable alternative to the deprecated functionality of NuSMV.

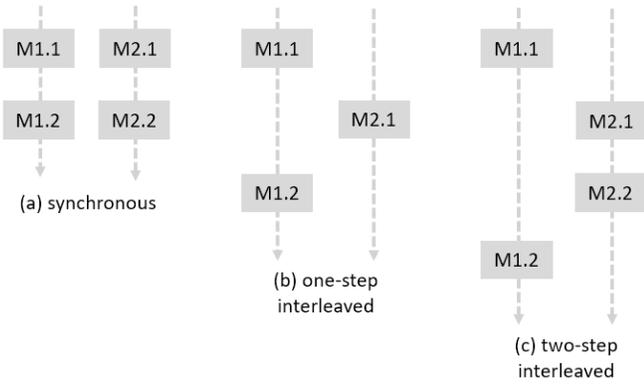

Fig. 6: Synchronous vs. Asynchronous Execution of Modules

## VI. Model Checking

We now had all specifications modeled and constraints formulated that needed to be satisfied. In this section, we outline our strategy to perform the model checking on this large-scale model, comprising ~1500 lines of code, 64 state variables, and 38 LTL constraints. Two of the 69 state variables were arrays allowing one of 42 possible failure states. Each single failure and double failure combination needed to be verified with a certain subset of the 38 constraints.

### A. Model Checking Strategy

The verification of all single and double failures posed a challenge in that regards, as the number of combinations add up to 42 + (42 x 42) = 1806 runs. Hence, we would need to automate this procedure instead of manually running the model checker. However, an essential prior step would be to perform a number of exemplary runs using selected double failure combinations. That way, we would be able to estimate the overall runtime and thereby assess the feasibility of our approach. Initial tests revealed that certain combinations of failures already resulted in exhaustive memory consumption or very long run-times, which we had to abort after several hours.

For more systematic testing and our actual verification, we designed an automation approach. The basic idea was to provide a template of the model that is enriched with a certain failure combination and subset of constraints for each run. A wrapper script or program would process this model template and enrich it with failure states and associated constraints from additional data files. Afterwards, the model checker could be started with the just generated model. With this kind of automation, we would be able to perform the 1806 verification runs in batch processing. We implemented this approach as a C++ program that we called *smvDriver*. We chose C++ for portability and maximum flexibility in extending the wrapper to support several parallel executions of NuSMV instances using multi-threading. The basic functionality of *smvDriver* is illustrated in Figure 7.

The program *smvDriver* is designed as a command line tool that can be started with arguments specifying the range of failures to be verified. As we stated above, there are 1806 double failure combinations resulting from a 42 x 42 matrix. E.g. running *smvDriver 1 1 2 2* would run four executions of the model checker, covering line 1, row 1 to line 2, row 2. For each execution, *smvDriver* performs the following steps: 1) Read the failures associated with the table row and column from the file *failures*; 2) Read the target operation mode from the matrix in file *target op-modes*; 3) Enrich the model template with the failure combination and associated constraints from the file *LTL specs*, and save the resulting file as *model instance*; 4) Run an instance of NuSMV by specifying the just created model instance as input. *smvDriver* then collects and aggregates the output of all NuSMV instances. When parsing the output, counterexamples are filtered out and saved per failure combination for easier analysis.

Using the tool *smvDriver*, we were able to perform several test runs with selected double failure combinations. It turned out that certain constraints could not be checked in a timely manner. Thus, we needed to consider further optimizations regarding the model, constraints, and the tool *smvDriver*.

> **Lesson Learned: Avoiding Past Temporal Operators**
>
> LTL provides past temporal operators to characterize properties of the path that led to the current state, e.g. *O p* evaluates to true, if condition *p* holds in one of the past time instants. We made heavy use of the *O* operator as the following simplified example shows:
>
> ```
> LTLSPEC G !(O OpModeA & O OpModeB)
> ```
>
> The formula ensures that `OpModeA` and `OpModeB` are never activated together in the same driving automation sequence. Using this operator frequently within LTL specifications caused unacceptable run-times. To address the issue, we tried a pre-evaluation of sub-expressions `O OpModeA` and `O OpModeB` in the model by creating two dedicated state variables `O_OpModeA` and `O_OpModeB`. We initialized them to *false* and set them to *true* once the respective fallback operation mode was activated. The following example shows this behavior for `OpModeA`:
>
> ```
> init (O_OpModeA) := false;
> next (O_OpModeA) := case
>     OpModeA : true;
>     true: O_OpModeA;
> esac;
> ```
>
> This workaround caused a flawless and quick evalu-ation of the affected LTL formula, hence we replaced all past temporal operators this way.

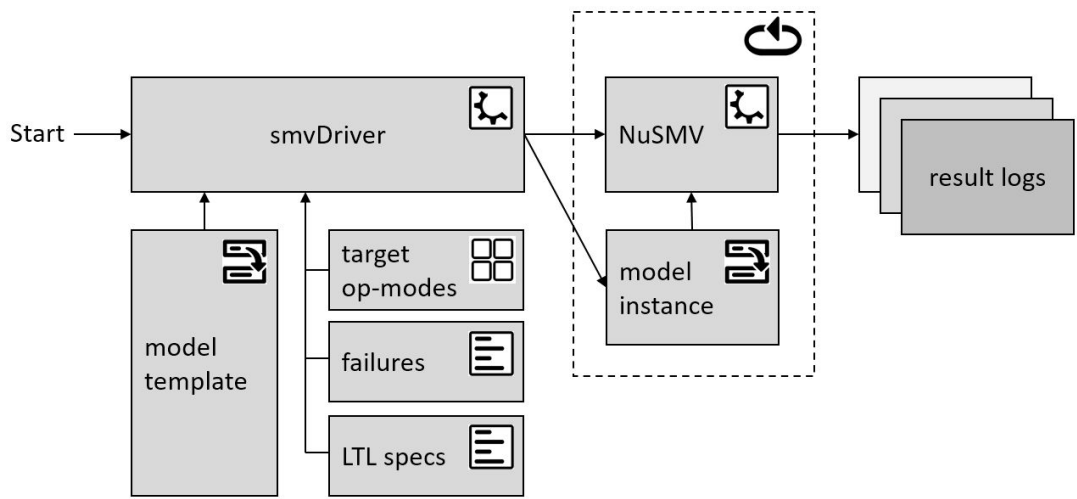

Fig. 7: Automation Strategy using the Tool smvDriver

**Lesson Learned: Multi-Core Processing for NuSMV**

Unfortunately, NuSMV does not natively support utilizing several processor cores. The solution would be to run several NuSMV instances in parallel to optimally utilize the available computing power. NuSMV allows to specify a single constraint to be checked by specifying its number as a command line argument. We could leverage this functionality to partition our verification runs. If we had 10 constraints to check in a model, we could run 10 instances of NuSMV with each instance checking a single constraint. Accordingly, we extended the functionality of *smvDriver* in the following way: Per *model instance*, several NuSMV instances are launched, one per LTL constraint. As a result, we were able to utilize as many processor cores as constraints included in the *model instance* and achieved a major speed-up. To fully utilize the processing power of our test machine's 40 cores, we eventually started several instances of *smvDriver* during our actual verification run.

### B. Bounded Model Checking

Various optimizations did not suffice our requirements on an acceptable run-time behavior. We tried NuSMV-specific optimizations like reducing the *Cone of Influence* through its command line argument *"-coi"*. This reduction *"forces the construction of a partial model which includes only those variables affected by the property being checked"* [30]. Its usage did not yield any improvements in our case, just as the argument *"-df"* to *"disable the computation of the set of reachable states"* [29]. It seemed that we reached the limits of NuSMV's BDD-based model checking.

Choi reported similar experiences when comparing NuSMV and SPIN model checkers for verifying flight guidance systems [13]. NuSMV's Bounded Model Checking [28] capability based on a Boolean Satisfiability (SAT) solver offered a possible solution. However, such solvers *"can prove the presence of errors in the model but cannot be used to prove the absence of errors, only the absence of errors reachable within k steps"* [31]. Reducing the exhaustive verification to a proof of error-absence within a certain number of steps is seemingly a major limitation. However, Biere considers this strategy as an *"enabling technology of many model checkers"* that is *"most relevant to industrial practice"* [28]. The question arose, if this technique is applicable in our scenario as well. As such, we will outline our basic considerations in this regard.

The initial run-up phase of the Vehicle Control System is finished after 15 steps. In failure-free operation, the system can be considered stable afterwards. The occurrence of failures causes the system to perform operation mode switches that have to be performed within a certain number of steps. The system can again be considered stable afterwards. If we would restrict the failure occurrence to a certain time span after the ramp-up phase, limiting the model checking to a certain number of steps would be a viable alternative. Clarke et al. elaborate on determining the correct bound [9]. They state that it can be derived from the *"diameter of the system, i.e., the least number of steps to reach all reachable states"* [9]. Unfortunately, they also state that finding this diameter is *"computationally hard"*. Instead, we determined an initial bound as shown in Figure 8, based on the considerations outlined above. That initial bound could be increased subsequently, given an acceptable run-time behavior.

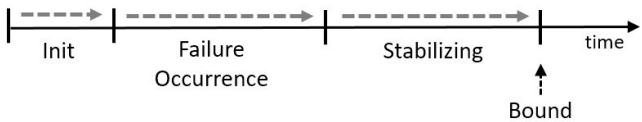

Fig. 8: Determining the Bound for BMC

Bounded Model Checking eventually brought the breakthrough. By setting the bound to 70 steps, each specification could be checked in less than 15 minutes, typically in 3 to 10 minutes. On our test machine accommodating four Intel Xeon Gold 5115 CPUs (with 10 cores each) and 256GB RAM, we were able to verify all 1806 failure combinations in ~30 hours. Memory consumption was modest compared to BDD-based model checking. Only few NuSMV instances consumed up to ~800MB which was not a limiting factor when processing 40 NuSMV instances in parallel.

## VII. Limitations

While this result can be regarded as a successful application of formal verification, several limitations need to be considered. We chose the symbolic model checker NuSMV for our verification. While its discrete time behavior makes the verification more tractable, it also means that the verification *"could miss faults that are caused by asynchronous execution order"* [1]. Considering alternative techniques and tools like SPIN, an explicit state model checker that supports asynchronous processes, could be a possible alternative. Choi reports on their experiences applying SPIN to *"complex synchronous mode logic with a large number of state variables, where SPIN is known to be not particularly efficient"* [13]. A possible solution could also bring hybrid systems verification tools that offer new ways to *"combine discrete and continuous behavior"* [32].

Employing Bounded Model Checking was necessary to achieve an acceptable runtime and overcome the *state explosion problem*. As a consequence, the chosen bound may cause the results to be unsound. However, we are confident that this issue could be addressed by setting significantly higher bounds or approximating the bound through one of the techniques mentioned by Clarke [9].

Finally, *"formal verification assumes the faults are in the design of a system"* and is as such able to find *"a violation of formal safety properties"*. The necessary abstraction in the course of modeling the system implies the omission of several aspects that could cause the real system to behave differently from the its model. We refer to our efforts made in order to abstract the communication between ECUs over different bus systems. Even considered faults of electronic components *"can cause behaviors that are unanticipated by the model"* [1]. A such, formal verification is a powerful tool to complement other testing and simulation techniques, but it cannot replace them.

## VIII. Conclusion

In this paper, we reported on our experiences gathered during model checking a Vehicle Control System. Starting from functional and safety specifications, we created a formal model of the architecture and its arbitration logic. We reported on our experiences during formalizing properties and deriving formula in temporal logic. The complexity of the resulting model required automating the verification procedure. For this, we designed an automation approach and developed an auxiliary program to partition and batch process the model checking. As a further optimization, it enables the model checker to utilize multiple processor cores. We explained our considerations on synchronous vs. asynchronous timing aspects and the decision for using NuSMV, a symbolic discrete time model checker. To achieve an acceptable runtime and overcome the *state explosion problem*, we used Bounded Model Checking that allowed our actual verification to complete in about 30 hours on a machine with 40 cores.

We shared several lessons learned tried to present our reasoning in a transparent and comprehensible way. As a final remark, we confirm the observations made by Choi already a decade ago [13]. He emphasizes the need for guidelines that support practitioners in *"choosing the right model checker for a particular problem domain"* [13]. Given the immense effort to model a complex system, wrong decisions made in the initial phase of a project can lead to costly false investments. We hope that architects, developers, and test engineers can leverage from our shared experiences in this regard to make more grounded decisions in similar projects.